\documentclass[11pt,twoside]{article}
\usepackage{asp2006}
\usepackage{epsf}
\usepackage{graphics}
\usepackage{lscape}
\usepackage{natbib}

\begin{document}
\title{Optical Variability of the Three Brightest Nearby
Quasars}
\author{C. Martin Gaskell, Andrew J. Benker, Jeffrey S. Campbell,\\
Thomas A. George, Cecelia H. Hedrick, Mary E. Hiller,
\\Elizabeth S. Klimek, Joseph P. Leonard, Shoji Masatoshi,
\\Bradley W. Peterson, Kelly C. Peterson, and Kelly M. Sanders}

\affil{Department of Physics \& Astronomy, University of Nebraska,
Lincoln, NE 68588-0111, USA}

\begin{abstract}

We report on the relative optical variability of the three brightest
nearby quasars, 3C 273, PDS 456, and PHL 1811. All three have
comparable absolute magnitudes, but PDS 456 and PHL 1811 are radio
quiet. PDS 456 is a broad-line object, but PHL 1811 could be
classified as a high-luminosity Narrow-Line Seyfert 1 (NLS1). Both
of the radio-quiet quasars show significant variability on a
timescale of a few days. The seasonal rms V-band variability
amplitudes of 3C 273 and PDS 456 are indistinguishable, and the
seasonal rms variability amplitude of PHL 1811 was only exceeded by
3C 273 once in 30 years of monitoring.  We find no evidence that the
optical variability of 3C 273 is greater than or more rapid than the
variability of the comparably-bright, radio-quiet quasars. This
suggests that not only do radio-loud and radio-quiet AGNs have
similar spectral energy distributions, but that the variability
mechanisms are also similar.  The optical variability of 3C 273 is
not dominated by a ``blazar'' component.

\end{abstract}

\section{Introduction}

    It is well known that all AGNs vary, but there are
many unanswered questions. Here we just address the question of
whether different types of AGN vary in the same manner. Two of the
obvious ways in which AGNs differ among themselves are:

(1) Radio-loudness and

(2) ``Eigenvector 1'' -- a correlated set of properties of which the
width of the broad lines is the best known (``Narrow-line Seyfert
1s'' = NLS1s, represent one extreme of eigenvector 1). Eigenvector 1
is widely believed to be a function of accretion rate.

Specific questions to consider are: (a) do bright radio-quiet AGNs
vary as much in the optical as radio-loud AGNs of comparable
brightness? and (b) do high-accretion-rate AGNs vary as much in the
optical as low-accretion-rate AGNs of comparable luminosity? The
conventional wisdom has been that radio-loud quasars show higher
amplitude optical variability than radio-quiet ones because there is
a contribution of a jet-related, non-thermal component in the
optical (a ``blazar'' component). There have been some indications
that NLS1s (widely considered to be high-accretion-rate AGNs) vary
less in the optical (see \citealt{Klimek06}), although NLS1s are
known to vary more in soft X-rays.

\section{Different AGNs have similar continuum shapes}

The spectral region showing the greatest and most rapid variability
is the hard to observe extreme ultraviolet (EUV) to soft X-ray
region. Because of the observational difficulties the flux in this
``EUV gap'' has to be extrapolated from the observed ultraviolet and
X-ray fluxes on either side of the gap. The observed continuum
shapes on both sides of the gap show considerable object-to-object
variations, and these differences are sufficiently large that they
have led many to postulate that there are fundamental differences in
how the energy is generated.  For example, \citet{Eracleous94}
suggested that radio galaxies might be powered by ion-supported
tori. However, the {\it observed} optical-UV continuum is heavily
influenced by intervening matter. In \citet{Gaskell04} and
\citet{GaskellBenker05} we demonstrate that the main differences in
the spectral-energy distributions (SEDs) of AGNs are consistent with
{\it reddening differences}. After allowance for reddening, we get
the surprising result that, at least for the AGNs we consider, AGN
SEDs from the near IR to $\sim 1200$ \AA \,{\it are
indistinguishable} for all classes of AGN and for all luminosities,
masses, and Eddington ratios.\footnote{it remains to be seen how
true this is for AGNs with the very lowest Eddington ratios.}

\subsection{The three brightest nearby quasars}

Given that the SEDs of all AGNs seem to be indistinguishable after
reddening corrections, the next question is: is the {\it
variability} also the same? To partially address this question we
have been looking at the variability of the three brightest nearby
quasars: 3C 273, PDS 456, and PHL 1811.

3C 273 ($z = 0.158$, $M_V = -26.6$), is the brightest, nearest, and
best-known, radio-loud quasar.  It has broad lines.  It is a
well-known variable and its light curve has been well studied for
decades. The other two quasars are more recently discovered, nearby,
radio-quiet quasars with luminosities comparable to 3C 273.  PDS 456
($z = 0.184$, $M_V = -26.9$) is a radio-quiet analog of 3C 273
\citep{Torres97, Simpson99}.  It has broad lines like 3C 273. PHL
1811 ($z = 0.192$, $M_V = -25.9$) is a narrow-line Seyfert 1 quasar
\citep{Leighly01}.  Figs. 1 and 2 show V-band light curves of PDS
456 and PHL 1811 from CCD observations were made with the 0.4-m
telescope at the University of Nebraska's Lincoln Observatory. The
observing and measurement procedures were as described in
\citet{Klimek04}.

%Fig. 1
\begin{figure}
\begin{center}
\epsfxsize = 110 mm \epsfbox{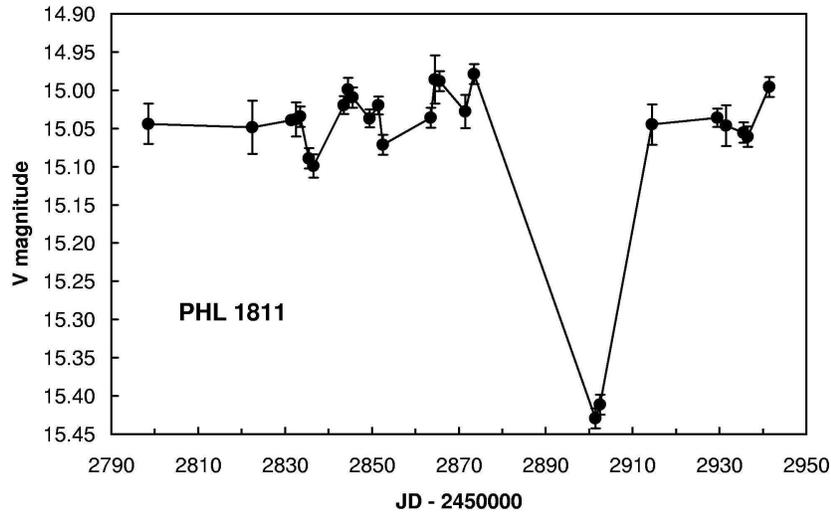} \caption{V-band
light curve for PHL 1811 in 2003. The lines connecting the points
are only to guide the eye.\label{PHL1811}}
\end{center}
\end{figure}

%Fig. 2
\begin{figure}
\begin{center}
\vspace{-0.2cm} \epsfxsize = 110 mm \epsfbox{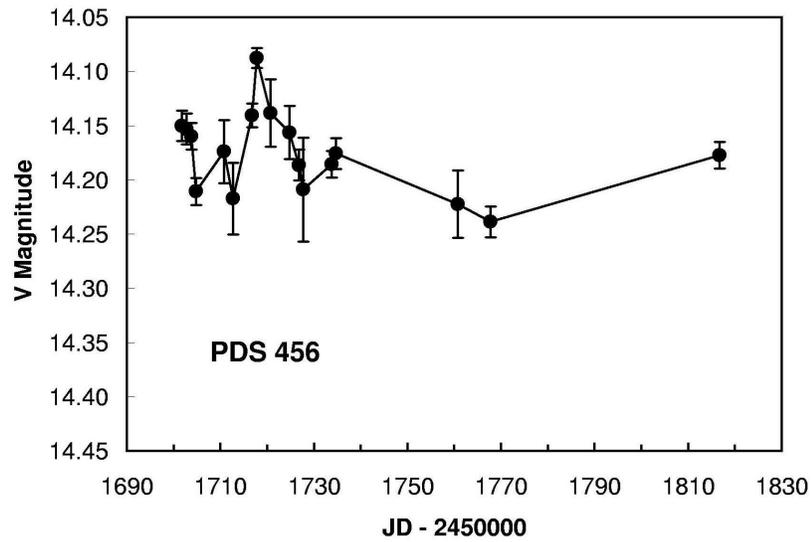}
\caption{V-band light curve for PDS 456 in 2000.\label{PDS456}}
\end{center}
\end{figure}

%\vspace{-0.5cm}

We can characterize the long-term variability by simply determining
the rms fluctuations for single observing seasons. For 3C~273 we
determined the rms fluctuations from 11 seasons of CCD photometry
given in \citet{Turler99}. We summarize our results in Table 1.

It can be seen from Table 1 that there is no evidence for the two,
bright, radio-quiet quasars to be less variable that the bright,
radio-loud quasar 3C 273. When considering the rms variability of
PHL 1811 it is interesting to note that, for 3C~273, the rms
seasonal variability has exceeded $\pm 0.100$ mag. only once over
the past 30 years. Even if we had not happened to catch the low
state of PHL 1811 around JD 2452895, its seasonal rms variation is
still similar to that of 3C 273.

\begin{center}
\begin{table}
%\begin{table}[b]
\caption{Seasonal RMS Variability.} \label{tab1}
\begin{tabular}{lcc}
\\
\hline
Object   &Year      &RMS (mag.)\\
\hline
PDS 456  &2000      &$\pm 0.042$\\
         &2003      &$\pm 0.036$\\
PHL 1811 &2003      &$\pm 0.109$\\
3C 273   &11 seasons&$\pm 0.042$\\
\hline
\end{tabular}
\end{table}
\end{center}

It is also interesting to compare short-term variability properties.
Although we did not often obtain daily sampling, for PDS 456 we did
find several $\sim 10$\% variations in the $V$-band over 24 hour
periods at the 99\% confidence level over three observing seasons
even without attempting to subtract out any host galaxy component.
PDS 456 is also an extreme X-ray variable with factor of two changes
in 8 hours \citep{Reeves02}. Thus, a bright, radio-quiet quasar {\it
can} show strong short-term variations in both the X-ray and optical
regions. PHL 1811 also shows significant variations over a day or
two, although the amplitude of short-term variability is not as
great for PHL 1811 as for PDS 456.

There is additional evidence that the short-term variability of
various classes of AGNs are similar.  Low-luminosity NLS1s show a
similar occurrence of optical sub-diurnal variability to non-NLS1
AGNs \citep{Klimek04, Klimek06}. And for radio-loud and radio-quiet
AGNs, \citet{Stalin04} found similar occurrences of optical
sub-diurnal variability.

\section{Conclusions}

Based on our photometry, it seems that, on both long and short
timescales, the conventional wisdom that radio-loud quasars should
be more variable is not correct, and there is no evidence that the
optical variability is fundamentally different for different classes
of AGNs.

\acknowledgements This research has been supported by the US
National Science Foundation through grant AST 03-07912, and also in
part by the Howard Hughes Foundation, Nebraska EPSCoR, the
University of Nebraska Layman Fund, and the University of Nebraska
Undergraduate Creative Activities and Research Experiences program.

\end{document}